\documentclass[conference]{IEEEtran}
\IEEEoverridecommandlockouts
\usepackage{cite}
\usepackage{amsmath,amssymb,amsfonts}
\usepackage{algorithmic}
\usepackage{graphicx}
\usepackage{textcomp}
\usepackage{xcolor}
\usepackage{bm}

\makeatletter
\IEEEsettextheight{0.81in}{1.03in}
\IEEEsettopmargin{t}{0.81in}
\IEEEquantizetextheight{c}
\IEEEsettopmargin{q}{0sp}
\makeatother

\newcommand{\modOp}[1]{\ \mathrm{mod}\ #1}

\def\BibTeX{{\rm B\kern-.05em{\sc i\kern-.025em b}\kern-.08em
    T\kern-.1667em\lower.7ex\hbox{E}\kern-.125emX}}
\begin{document}

\title{Radio Stripe-Based Distributed ISAC System with Dynamic Sensing-Communication Reconfiguration\\
\thanks{This work is partially supported by the Research Council of Finland (Grants 362782 (ECO-LITE), and 369116 (6G Flagship)).}
}

\author{\IEEEauthorblockN{1\textsuperscript{st} Osmel Mart\'{i}nez Rosabal}
\IEEEauthorblockA{\textit{Centre for Wireless Communications} \\
\textit{University of Oulu}\\
Oulu, Finland \\
osmel.martinezrosabal@oulu.fi}
\and
\IEEEauthorblockN{2\textsuperscript{nd} Onel Alcaraz L\'{o}pez}
\IEEEauthorblockA{\textit{Centre for Wireless Communications} \\
\textit{University of Oulu}\\
Oulu, Finland \\
onel.alcarazlopez@oulu.fi}
}

\maketitle

\begin{abstract}
Integrated sensing and communications (ISAC) has emerged as an intrinsic service of upcoming 6G wireless systems, enabling the reuse of communication signals for environmental sensing and supporting context-aware network functionalities. Meanwhile, the evolution of the wireless infrastructure toward distributed systems creates new opportunities for collaborative sensing from spatially separated nodes. Motivated by this trend, this work investigates a radio stripe aided ISAC system as a low-complexity implementation of a distributed system. We study the trade-off between achievable sum rate and sensing precision when downlink signals are used for target localization within the service area. By exploiting the architectural homogeneity of the radio stripes transceivers, each unit can be dynamically configured to operate in either communication or sensing mode. We formulate a targets localization problem considering the measurements of multiple sensing-communication configurations. Due to the large number of measurements and the continuity of the search space, we propose discretizing the service are and then solve the estimation problem in batches. The targets are finally estimated using a fusion strategy. Our results show that increasing the number devices and sensing APUs boosts sensing precision at the expense of degrading the sum rate. The latter remains constant for a given number of communication APUs regardless of their positions. Moreover, changing the number of antennas reveals a non-monotonic impact on sensing performance due to the trade-off between array gain and illumination uniformity.
\end{abstract}


\IEEEPARstart{T}{he} evolution toward the sixth-generation of mobile systems is leaning towards service convergence, the native integration of artificial intelligence, and the design of inherently resilient and sustainable networks \cite{Alves.2025,Lopez.2023,Bariah.2024}. The integration of communication and sensing (ISAC) has emerged as a key enabler of this paradigm shift, bridging conventional radar and communication operations within a unified framework. This capability allows the network to actively perceive and learn from its environment using its own infrastructure to improve not only communication performance but to enable context-aware services for sustainable and resilient networks and society \cite{Liu.2022}. The impact of ISAC materializes across a wide range of tightly coupled communication-sensing applications, including vehicle-to-everything, smart home, smart manufacturing, remote sensing, environmental monitoring, and human-computer interaction \cite{Liu.2022}.

In this context, the distributed nature of modern wireless infrastructures, which was driven by the need to improve coverage, reliability, and spectral efficiency, emerges as a key architectural feature to enhance ISAC services. By leveraging geographically separated nodes, the network can operate as a spatially distributed sensing aperture, providing multiple transmit-receive viewpoints of the environment. This principle, long recognized in distributed radar systems \cite{Haimovich.2008} and recently validated in practical massive multiple-input and multiple-output (MIMO) testbeds \cite{Sakhnini.2022}, yields increased spatial diversity that enriches delay, angle, and Doppler information. As a result, the sensing problem becomes better conditioned, improving resolution, robustness to blockage, and spatial coverage compared to mono-static configurations \cite{Chen.2024}. These benefits are particularly pronounced in dynamic environments, where the most informative sensing perspectives vary over time and can be opportunistically exploited through distributed ISAC.

Notably, current infrastructure is primarily deployed for communication purposes, with no sensing resources provisioned by default. This limitation has motivated the repurposing of existing communication signals to enable sensing in recent works \cite{Tagliaferri.2025,Murtada.2025,Li.2024,Murtada.2023,Leyva.2025}. For instance, the synergistic communication and sensing frameworks proposed in \cite{Murtada.2025} exploit downlink transmissions to jointly enable data delivery and distributed scene reconstruction. Therein, the authors build on the concept of virtual users introduced in \cite{Li.2024} as a tool to enhance illumination in zones where no users are deployed. Moreover, \cite{Murtada.2023} presents a novel approach to reconstruct a unique image of an observed environment by widely distributed radar sensors. They consider a mono-static configuration where each radar receives the reflections due to its own signals and does not process the reflections induced by transmissions from others. 

While the advantages of repurposing the current infrastructure are clear, communication-sensing co-design strategies have shown tremendous potential. For instance, \cite{Tagliaferri.2025} proposes a joint design coupling waveform, precoding, power allocation and nodes transmitting and receiving roles in a distributed MIMO system. Notably, the results presented therein highlight the need of the phase-coherent operation among the nodes. However, such coordination also raises scalability challenges, particularly in terms of fronthaul signalling overhead, since signal processing algorithm run in a distributed manner. To address this issue, \cite{Leyva.2025} propose a two-stage beamforming design to maximize the users' sum signal-to-interference-plus-noise ratio subject to transmitter power budget and a minimum sensing signal-to-noise ratio (SNR).

The aforementioned advantages open up significant opportunities for distributed antenna systems implemented via radio stripes. A radio stripe consists of a large number of distributed antenna processing units (APUs) embedded along physical structures such as walls or ceilings and interconnected through a shared fronthaul infrastructure \cite{Lopez.2023}. Radio stripes enable low-complexity distributed MIMO implementations and almost imperceptible installation, which is critical in many use cases. Moreover, radio stripes enable a phase-coherent, spatially distributed aperture that improves ISAC capability beyond conventional distributed or co-located arrays \cite{Fascista.2025}.

Motivated by these advantages this work proposes a radio stripes based multi-static ISAC system. Specifically, we consider that APUs are assigned to either communication or sensing roles. We consider that devices are served via joint coherent transmission from all communication APUs over dedicated subcarriers. Different from previous works, we account for the reflected signals induced by the communication transmissions from different APUs. Our contributions are: i) we leverage the homogeneity of the APUs to dynamically change their roles to boost estimate of the targets location; ii) we propose discretizing the service area to obtained partial estimates by each communication-sensing configuration and then fuse them to tackle the problem complexity; iii) we illustrate how the sensing precision enhances with the number of sensing APUs and served devices at the cost of degrading the data rate when the number of communication APUs is reduced; iv) we show that, for a given configuration, varying the positions of the sensing APUs improves sensing precision while having a negligible impact on the achievable data rate; and v) we illustrate that sensing precision exhibits a non-monotonic dependence on the number of antennas due to a trade-off between array gain and illumination homogeneity.

\textbf{Outline:} Next, Section~\ref{sec:systemModel} introduces the system model and problem formulation. Section~\ref{sec:proposedSolution} proposes a discretized sensing model and the recovery of a composite target representation. Subsequently, Section~\ref{sec:numericalResults} analyzes the sensing-rate trade-offs numerically. Finally, Section~\ref{sec:conclusions} summarizes the findings and concludes this work. 

\textbf{Notation:} We use boldface lowercase letters to denote column vectors and boldface uppercase letters to denote matrices. $\lVert \mathbf{x} \rVert_p$ denotes the $\ell_p$-norm of $\mathbf{x}$, while $(\cdot)^T$, $(\cdot)^H$, $(\cdot)^*$, $|\cdot|$, $\lfloor \cdot \rfloor$, and $\mathrm{mod}(\cdot)$, denote the transpose, Hermitian transpose, complex conjugate, absolute value, floor, and modulus operation, respectively. Also, $\mathbb{C}$ is the set of complex numbers with imaginary unit $j = \sqrt{-1}$, and $\mathbf{I}_{N \times M}$ denote an $N \times M$-dimensional identity matrix. 

\section{System model and problem formulation}\label{sec:systemModel}
\begin{figure}
    \centering
    \includegraphics[width=\linewidth]{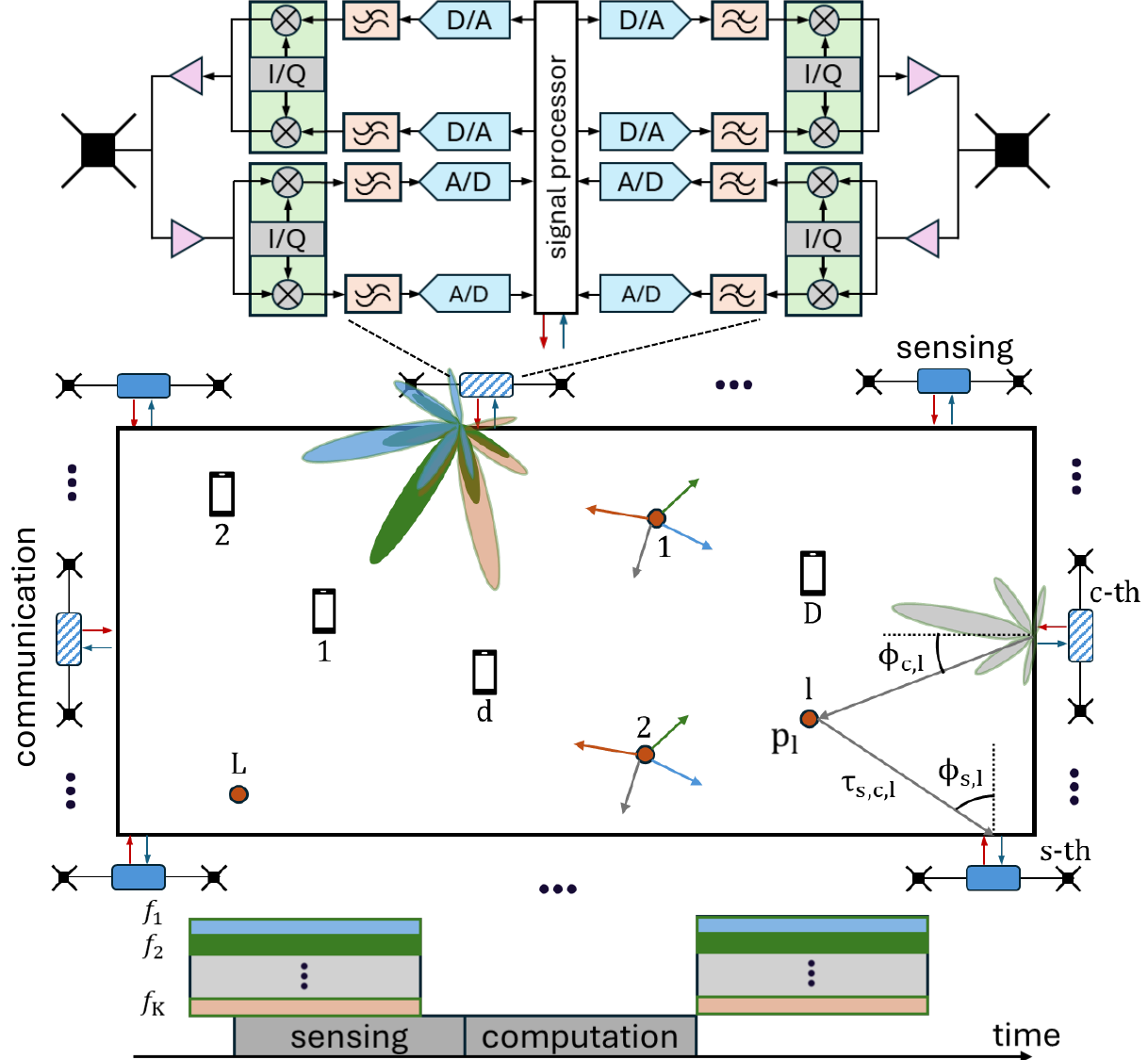}
    \caption{Radio stripes-aided ISAC system where APUs are assigned to communication or sensing roles. Communication APUs serve devices in a 2D service area using OFDMA.}
    \label{fig:systemModel}
\end{figure}
Consider the radio stripe system depicted in Fig.~\ref{fig:systemModel}, which serves $D$ single-antenna communication devices distributed on a 2D service area. This geometry assumes that the devices are approximately confined to a common plane which is typical in indoor deployments \cite{Murtada.2025}. The radio stripe comprises multiple APUs, each equipped with $M$ antenna elements, which share a common serial bus for synchronization, data transfer, and power supply. The APUs are dynamically partitioned into two disjoint sets of $C$ data communication and $S$ sensing APUs, respectively. Data communication APUs serve the devices via orthogonal frequency-division multiple access (OFDMA), whereas the sensing APUs acquire/process the reflected signals induced by the communication transmissions.

Each user is assigned a unique subcarrier among the $K > D$ in the OFDMA bandwidth and served via joint coherent transmission from all communication APUs. Considering that a common symbol is transmitted from all communication APUs, the transmitted signal on the $k$-th subcarrier is
\begin{equation}
    \mathbf{x}_k = \mathrm{vec}\Big( \big[\mathbf{u}_{1,k}, \ldots, \mathbf{u}_{c,k}, \ldots, \mathbf{u}_{C,k},\big] \Big) d_k,
\end{equation}
where $\mathbf{u}_{c,k} \in \mathbb{C}^{M \times 1}$ is the digital precoder employed by the $c$-th communication APU on the $k$-th subcarrier, with transmit power $\lVert \mathbf{u}_{c,k} \rVert^2_2 = p_{c,k}$, and $d_k$ is a data symbol, assumed independent across subcarriers and satisfying $\mathbb{E}[d_k d^*_{k'}] = 1$ for $k = k'$ and $0$ otherwise. Then the received signal is 
\begin{equation}
    y_k = \mathbf{h}^H_k \mathbf{x}_k + w_k,
\end{equation}
where $\mathbf{h}_k \in \mathbb{C}^{CM \times 1}$ denotes the complex channel coefficients to the device allocated to the $k$-th subcarrier computed as
\begin{equation}
    \mathbf{h}_k = \mathrm{vec}\Big( \big[\mathbf{a}_k(\phi_{1,u}), \ldots, \mathbf{a}_k(\phi_{c,u}), \ldots, \mathbf{a}_k(\phi_{C,u})\big] \Big),
\end{equation}
and $w_k \sim \mathcal{CN}(0,\sigma^2)$ is the additive circularly symmetric Gaussian noise with effective power $\sigma^2$. For simplicity, we assume homogeneous (not distance-dependent) path-loss and incorporate its effect into $\sigma^2$ (cf. Section IV), without loss of generality \cite{Li.2023,Li.2024,Murtada.2025}. Moreover,
\begin{equation}
    \mathbf{a}_k(\phi) = 
        \left[ 1,\ e^{j2\pi \frac{f_k}{f_c}\Delta \sin{\phi}}, \ldots, e^{j2\pi \frac{f_k}{f_c}\Delta (M-1) \sin{\phi}} \right]^T,
\end{equation}
is the APU array steering vector at the subcarrier frequency $f_k$ and $\Delta$ is the inter-antenna separation normalized with respect to the wavelength at the carrier frequency $f_c$. Considering coherent maximum ration transmission the signal-to-noise (SNR) ratio at the $k$-th subcarrier is
\begin{equation}
    \Gamma_k = \sum_{c=1}^{C} \frac{|\mathbf{a}^H_{k}(\phi_{c,u}) \mathbf{u}_{c,k}|^2}{\sigma^2} = \sum_{c=1}^{C} \frac{p_{c,k}\lVert\mathbf{a}_{k}(\phi_{c,u})\rVert^2_2}{M\sigma^2}.
\end{equation}
The service area contains $L$ targets whose responses capture the relevant components of the environment reflectivity. Diffuse and background scattering are suppressed through preprocessing and are not explicitly modeled, as in \cite{Murtada.2023,Murtada.2025}. Hence, the signal received at the $s$-th sensing APUs in the subcarrier $k$ is 
\begin{equation}
    \mathbf{y}_{s,k} = \mathbf{H}_{s,k}\mathbf{x}_k + \mathbf{w}_{s,k},
\end{equation}
where $\mathbf{w}_{s,k} \sim \mathcal{CN}(0,\sigma^2\mathbf{I}_{M\times 1})$ is the additive circularly symmetric Gaussian, again with effective power $\sigma^2$, and 
\begin{equation}
    \mathbf{H}_{s,k} = \left[\mathbf{H}_{s,1,k}, \ldots, \mathbf{H}_{s,c,k}, \ldots, \mathbf{H}_{s,C,k} \right] \in \mathbb{C}^{M \times CM},
\end{equation}
is the matrix of complex channel coefficients with sub-matrices
\begin{equation}
    \mathbf{H}_{s,c,k} = \sum_{l=1}^{L} z_l e^{-j2\pi f_k \tau_{s,c,l}} \mathbf{a}_k(\phi_{s,l}) \mathbf{a}_k(\phi_{c,l})^H, 
\end{equation}
where $\phi_{s,l}$ and $\phi_{c,l}$ are the azimuth angles of the $l$-th target with respect to the $s$-th sensing and $c$-th communication APUs, respectively. Furthermore, $\tau_{s,c,l}$ denotes the propagation delay from the $c$-th communication APU to the $l$-th scatterer and subsequently to the $s$-th sensing APU. These quantities are determined by the position of the $l$-th target, denoted by $\mathbf{b}_l \in \mathbb{R}^2$, and the locations of the corresponding APUs. 
% Besides, the vector $\mathbf{z} \in \mathbb{C}^{L \times 1}$, with entries $\{z_l\}$, collects the radar cross sections of the targets. In this work, the magnitudes $\{|z_l|\}$ are primarily used to indicate the presence of a target at a given location, rather than to characterize its absolute reflectivity across different illumination angles. 

 Now, for a given pair $(S,C)$, different assignments of sensing and communication roles lead to distinct illumination and observation patterns over the service area. For instance, when each side of the service area hosts $2$ APUs, the total number of APUs is $8$. Assigning exactly $2$ APUs to sensing roles yields $\binom{8}{2} = 28$ distinct role configurations. Each observation corresponds to the signal measured at a sensing APU, determined by its location and the specific set of communication APUs illuminating the region which yields $28 \times 2 = 56$ observations in total.  Let us denote the resulting collection of sensing observations by $\mathcal{S}$. Our goal is to recover the underlying scene representation, characterized by the locations and reflectivities of the $L$ targets, from the collection of measurements acquired across all sensing APUs.
\begin{subequations}\label{P1}
    \begin{alignat}{2}
    \mathbf{P1:} &\underset{\mathbf{z}_l, \mathbf{b}_l}{\mathrm{min.}} \ && \sum_{s \in \mathcal{S}} \lVert \mathbf{y}_{s,k} - \mathbf{H}_{s,k} \mathbf{x}_{s,k} \rVert^2_2. \label{P1a}
    \end{alignat}
\end{subequations}
Note that $\mathbf{P1}$ is non-convex due to the nonlinear dependence of the sensing matrices $\mathbf{H}_{s,k}$ on the target positions $\{\mathbf{b}_l\}$, which couple the propagation delays and angles across all sensing observations. In addition, the search space is continuous and grows rapidly with the number of targets, as it involves all possible spatial realizations of the targets and their associated reflectivities. This complexity is further increased by the large number of observations induced by different sensing and communication APU role assignments, each giving rise to a distinct observation, thereby rendering the problem computationally infeasible.

\section{Consensus-based reconstruction with distributed fusion}\label{sec:proposedSolution}
In this section, we rely on a consensus image recovery for given assignments of sensing and communication roles with a final fusion step among all possible observations. 

\subsection{Problem reformulation}\label{subsec:problemReformulation}
We begin by discretizing the service are with a set $\mathcal{I}$ of $I$ points. Each point $i \in \mathcal{I}$ is associated with Cartesian coordinates given by
\begin{align}
    \left(\left\lfloor \frac{(i-1)}{\sqrt{I}} \right\rfloor - 1, ((i-1) \modOp{\sqrt{I}})+1\right)\delta,
\end{align}
where $\delta$ denotes the grid spacing in both directions. Moreover, we focus on a particular assignment $n$, with corresponding observation set $\mathcal{S}_n \subset \mathcal{S}$ among $N$ possible ones. For this assignment, the discretized $c$-th sub-matrix of the channel $\mathbf{H}_{s,k}$ (with a slight abuse of notation) can be computed as
\begin{equation}
    \mathbf{H}_{s,c,k} = \sum_{i=1}^{I} z_{s,i} e^{-j2\pi f_k \tau_{s,c,i}} \mathbf{a}_k(\phi_{s,i}) \mathbf{a}_k(\phi_{c,i})^H, 
\end{equation}
where $\mathbf{z}_s \in \mathbb{C}^{I \times 1}$ is a vector whose entries $z_{s,i}$ represent the radar cross section associated with the $i$-th spatial point as observed by the $s$-th sensing APU. In this work, the magnitudes $\{|z_{s,l}|\}$ are primarily used to indicate the presence of a target at a given location, rather than to characterize its absolute reflectivity across different illumination angles. Accordingly, we drop the dependence on the communication APU index $c$.

If the grid is sufficiently fine, the true scatterers occupy only a small subset of these locations, meaning that the channel representation becomes sparse. This observation motivates the use of compressed sensing \cite{Li.2023} to obtain position of the scatterers, for which we first construct an appropriate sensing matrix. We begin by rewriting
\begin{equation}
    \mathbf{H}_{s,k}\mathbf{x}_k = \sum_{c=1}^{C} \mathbf{H}_{s,c,k}\mathbf{u}_{c,k}d_k = \sum_{i=1}^I z_{s,i} \Phi_{s,k,i}, 
\end{equation}
where the $i$-th column of the sensing matrix $\Phi_{s,k} \in \mathbb{C}^{\times I}$ is
\begin{equation}
    \Phi_{s,k,i} = \sum_{c=1}^C e^{-j2\pi f_k \tau_{s,c,i}} \mathbf{a}_k(\phi_{s,i}) \mathbf{a}_k(\phi_{c,i})^H \mathbf{u}_{c,k}d_k.
\end{equation}

Then, we can re-write the received signal in vector form as
\begin{equation}
    \mathbf{y}_{s,k} = \bm{\Phi}_{s,k}\mathbf{z}_{s} + \mathbf{w}_{s,k}.
\end{equation}
and, by stacking the signals across all subcarriers, the aggregated observation at the $s$-th APU is computed as
\begin{equation}
    \mathbf{y}_s = \bm{\Phi}_s \mathbf{z}_s + \mathbf{w}_s.
\end{equation}

Our objective is to recover a composite target representation by jointly exploiting the measurements collected across all sensing APUs, while enforcing a shared global structure across local estimates. This leads to the following consensus optimization problem \cite{Murtada.2023}
\begin{subequations}\label{P2}
    \begin{alignat}{2}
    \mathbf{P2:} &\underset{\{\mathbf{z}_s\}, \mathbf{z}_G}{\mathrm{min.}} \ && \sum_{s=1}^{S} \lVert \mathbf{y}_s - \bm{\Phi}_s \mathbf{z}_s \rVert^2_2 + \alpha \lVert \mathbf{z}_G \rVert_1, \label{P2a}\\
    &\text{s.t.} \ && \mathbf{z}_s - \mathbf{z}_G = \mathbf{0}, \forall s \in \mathcal{S}, \label{P2b}
    \end{alignat}
\end{subequations}
where $\mathbf{z}_G \in \mathbb{C}^{I \times 1}$ represents the magnitude of the global image and $\alpha > 0$ enforces solution sparsity. 

\subsection{Proposed solution}
To solve $\mathbf{P2}$, we adopt the consensus alternating direction method of multipliers (ADMM) framework in \cite[Section III.A]{Murtada.2023}. Introducing dual variables $\bm{\gamma}_s$ and a penalty parameter $\beta > 0$, the augmented Lagrangian is given by 
\begin{align}
    \mathcal{L}(\{\mathbf{z}_s\}, \mathbf{z}_G,\bm{\gamma}_s) = \sum_{s=1}^{S} & \Big[ \,
    \lVert \mathbf{y}_s - \bm{\Phi}_s \mathbf{z}_s \rVert_2^2
    + \bm{\gamma}_s^H(\mathbf{z}_s - \mathbf{z}_G) \nonumber \\
    & + \frac{\beta}{2} \lVert \mathbf{z}_s - \mathbf{z}_G \rVert_2^2 
    \Big]
    + \alpha \lVert \mathbf{z}_G \rVert_1.
\end{align}

Then, at each iteration $t$ the local measurements of each APU are updated as
\begin{align}
    \mathbf{z}^{(t+1)}_s = (\bm{\Phi}^H_s\bm{\Phi}_s + \beta \mathbf{I})^{-1}(\bm{\Phi}^H_s\mathbf{y}_s + \beta \mathbf{z}^{(t)}_G - \bm{\gamma}_s^{(t)}).
\end{align}
Meanwhile, the update of the global measurements vector can be obtained by solving 
\begin{equation}\label{eq:globalUpdate}
    \begin{aligned}
        \mathbf{z}^{(t+1)}_G &= \arg\min_{\mathbf{z}_G} \quad \alpha \lVert \mathbf{z}_G \rVert_1 + \frac{S\beta}{2}\lVert \mathbf{z}_G \rVert^2_2 \\
        & + \sum_{s=1}^{S} \left(\bm{\gamma}^{(t)}_s - \beta \mathbf{z}^{(t+1)}_s \right)^H  \mathbf{z}_G + \frac{\beta}{2} \sum_{s=1}^S \lVert \mathbf{z}^{(t+1)}_s \rVert^2_2,
    \end{aligned}
\end{equation}
whose objective function is a rearrangement of the Lagrangian for given $\{\mathbf{z}^{(t+1)}_s\}$ and $\bm{\gamma}^{(t)}_s$. By completing the perfect square, \eqref{eq:globalUpdate} reduces to the equivalent formulation (up to a constant term)
\begin{equation}\label{eq:globalUpdateProximal}
    \mathbf{z}^{(t+1)}_G = \underset{\mathbf{z}_G}{\arg\min} \ \alpha \lVert \mathbf{z}_G \rVert_1 + \frac{S\beta}{2} \lVert \mathbf{z}_G - \mathbf{z}^{(t)} \rVert^2_2
\end{equation}
where 
\begin{equation}
    \mathbf{z}^{(t)} = \frac{1}{S}\sum_{s=1}^S \mathbf{z}^{(t+1)}_s - \frac{1}{\beta S}\sum_{s=1}^S \bm{\gamma}^{(t)}_s.
\end{equation}
The solution to \eqref{eq:globalUpdate} corresponds to the proximal operator of the $\ell_1$-norm, yielding the soft-thresholding update \cite{Parikh.2014}
\begin{equation}
    \mathbf{z}^{(t+1)}_G = \mathcal{S}_{\alpha/\beta S} (\mathbf{z}^{(t)}),
\end{equation}
where the $i-$th entry is
\begin{equation}
    \left[\mathcal{S}_{\alpha/\beta S}\right]_i = 
    \begin{cases}
        \operatorname{sign}(z^{(t)}_i)\,\left(|z^{(t)}_i| - \frac{\alpha}{\beta S}\right), & |z^{(t)}_i| > \frac{\alpha}{\beta S}, \\
        0, & |z^{(t)}_i| \le \frac{\alpha}{\beta S}.
    \end{cases}
\end{equation}
Finally, the dual variable is updated as
\begin{align}
    \bm{\gamma}^{(t+1)}_s = \bm{\gamma}^{(t)}_s + \beta \left(\mathbf{z}^{(t+1)}_s - \mathbf{z}^{(t+1)}_G\right).
\end{align}
The consensus ADMM framework iteratively updates the variables until convergence. Although convergence is assessed via primal and dual residuals satisfying predefined tolerances in \cite{Murtada.2023}, in this work we employ a fixed number of iterations to ensure a controlled computational complexity.

Consensus ADMM is executed independently for each assignment $n$, yielding a set of global estimates $\{\mathbf{z}^{(n)}_{G}\}$, which are finally fused to obtain a refined global estimation. Specifically, we adopt a weighted aggregation strategy in which the normalized global estimates are combined using weights derived from the normalized primal residuals
\begin{equation}
    \mathbf{\Tilde{z}}_G = \sum_{n=1}^{N} r_n \frac{\mathbf{z}^{(n)}_{G}}{\lVert\mathbf{z}^{(n)}_{G}\rVert_2}.
\end{equation}

\section{Numerical results}\label{sec:numericalResults}
In this section we evaluate the performance of the system. The OFDMA bandwidth is centred at $f_c = 5.955~$GHz and comprises $K = 64$ subcarriers spaced by $312.5~$kHz \cite{ETSI.2023}. The antennas at each APU are spaced at half the wavelength corresponding to the carrier frequency $f_c$. Moreover, we consider that there are $L = 10$ scatterers in service area which is discretized using $I = 400$ points. The considered region has a square shape with a perimeter of $240~\mathrm{m}$, and the same number of APUs is installed along each side. The transmission power budget is $1~$W and the effective noise power is $\sigma^2 = 10^{-6}$, hence including also path-losses in the order of $90~$dB.. 

We rely on precision to evaluate targets detection performance, which for any estimate $\mathbf{\Tilde{z}}_G$ (fused or not) is defined as
\begin{equation}
    \mathrm{precision} = \frac{\sum_{i=1}^{I} \mathbf{1}\{z_{G,i} > 0\} \mathbf{1}\{\Tilde{z}_{G,i} > 0.85\}}{\sum_{i=1}^{I} \mathbf{1}\{z_{G,i} > 0\}},
\end{equation}
$0.85$ is used as the detection threshold.

For given values of $S$ and $C$, we evaluate the performance of $N = \binom{S + C}{C}$ possible configurations and average out the results over $1000$ uniform devices deployments. Furthermore, we set $\alpha = 1.8$, $\beta = 100$, and run consensus-ADMM for $50$ iterations. These parameters are empirically selected based on reconstruction performance and convergence.

\begin{figure}[t]
    \centering
    \includegraphics[width=\linewidth]{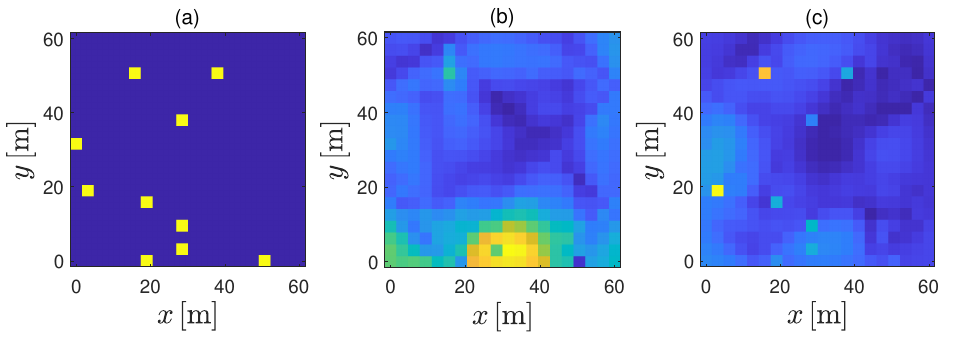}
    \caption{Non-fused scene reconstruction quality for $S = C = 4$ in specific APU roles assignment and different number of served devices: (a) original, (b) $D = 6$, (c) $D = 22$.}
    \label{fig:scenes}
\end{figure}

\begin{figure}[t]
    \centering
    \includegraphics[width=\linewidth]{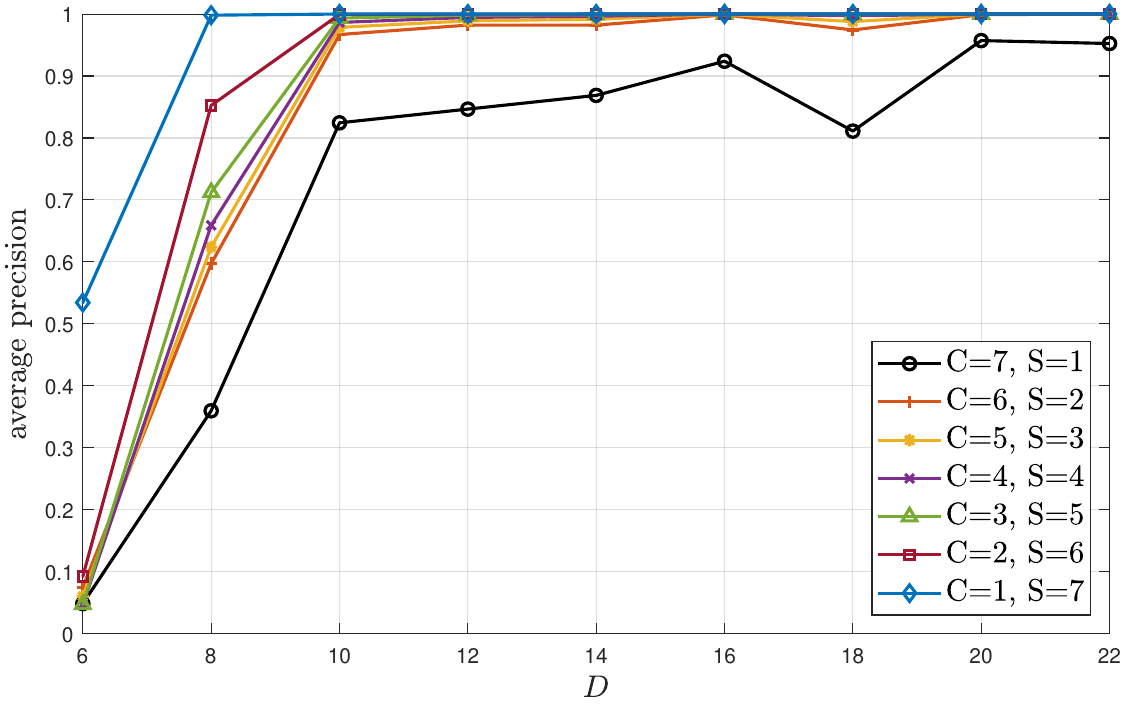}
    \caption{Average precision in the fused image vs $D$ for $M = 4$.}
    \label{fig:resultsVSU}
\end{figure}

\begin{figure}[t]
    \centering
    \includegraphics[width=\linewidth]{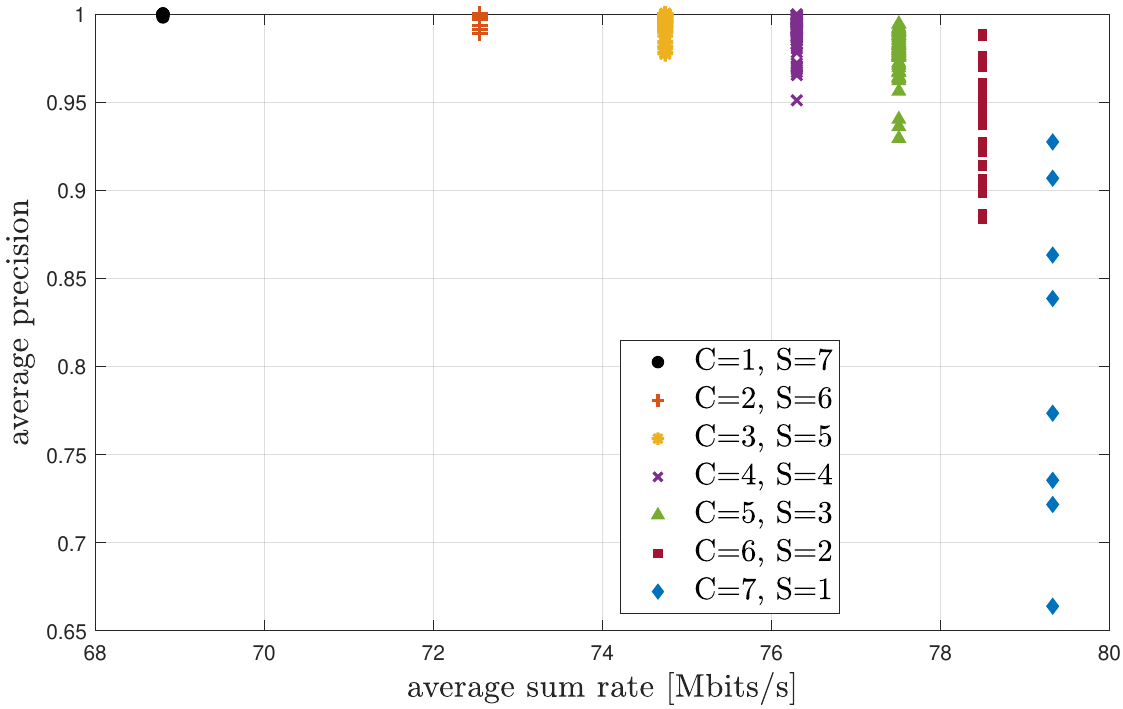}
    \caption{Average precision vs average sum rate for different APU configurations, $D = 12$, and $M = 4$.}
    \label{fig:pareto}
\end{figure}

Fig.~\ref{fig:scenes} illustrates how the reconstruction quality improves sharply as the number of devices increases. This is because the communication beams are more likely to illuminate homogeneously the service area from different positions. 
Consequently, as Fig.~\ref{fig:resultsVSU} illustrates, the fused reconstruction of the environment reflectivity is boosted, specially when simultaneously increasing the number of sensing APUs. This, however, degrades the achievable rate by approximately $0.8773~$Mbit/s from the best to the worst configuration. The inherent trade-offs between communication and sensing capabilities in ISAC systems are better shown in Fig.~\ref{fig:pareto} which illustrates the average precision vs average sum rate results. Interestingly, modifying the assignment of communication roles for given a $S$ and $C$ does not affect the achievable sum rate due to the distributed nature of the network; however, it does influence the sensing precision, particularly in configurations with a fewer sensing APUs.

\begin{figure}[t]
    \centering
    \includegraphics[width=\linewidth]{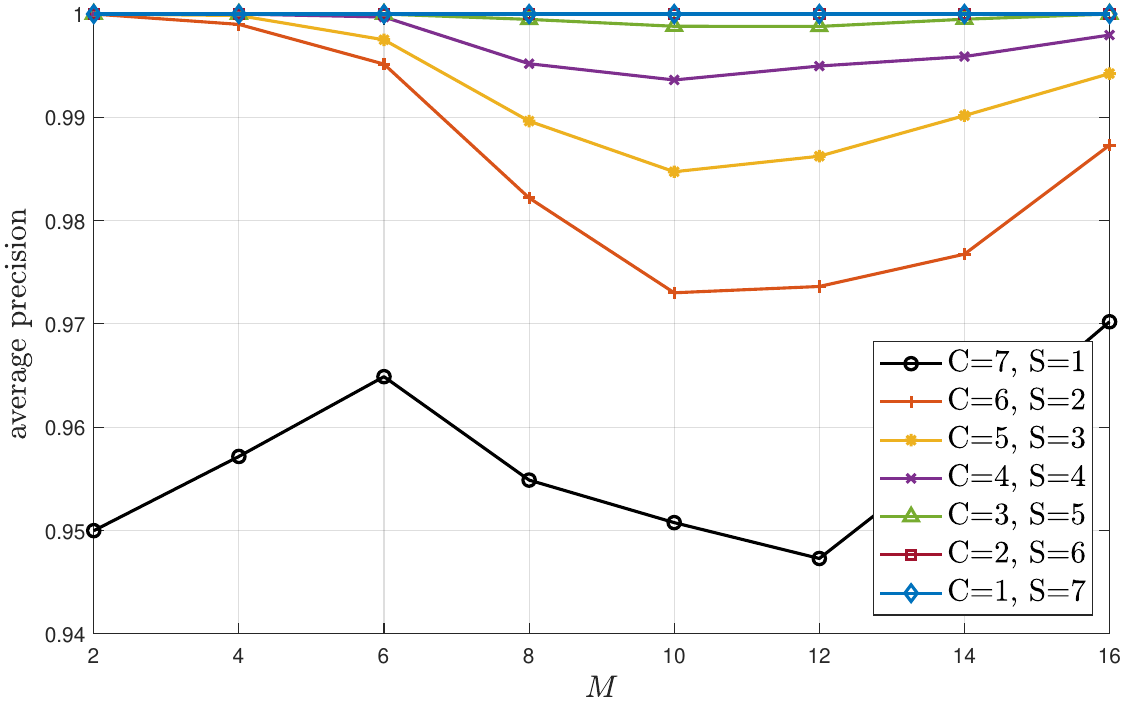}
    \caption{Average precision in the fused image vs $M$ for $D = 20$.}
    \label{fig:resultsVSM}
\end{figure}

Finally, Fig.~\ref{fig:resultsVSM} shows a non-monotonic behavior of the average precision as a function of the $M$ for different configurations. Increasing $M$ initially enhances sensing performance due to improved array gain and angular resolution. However, at intermediate values of $M$, the increased beam directivity concentrates energy toward specific directions, reducing the uniformity of illumination across the region and degrading reconstruction quality. For larger $M$, the higher array gain compensates for this loss of coverage, resulting in a recovery of the sensing performance. Such general trend is less pronounced in configurations with larger number of sensing APUs due to the increased spatial diversity and redundancy of observations.

\section{Conclusions}\label{sec:conclusions}
This work investigated a radio stripes based multi-static ISAC system operating under phase coherence. We formulated the scene reconstruction problem under the constraint that each APU is assigned either a communication or a sensing function and that the network is allowed to dynamically reassign these roles across APUs, yielding multiple observations. To address this problem, we employed a consensus-based reconstruction scheme followed by a fusion stage that combines estimates obtained from different communication and sensing configurations. This approach is carried over a discretized version of the service area and allows the network to observe the service area from multiple spatial viewpoints. The results showed that sensing precision improves as the number of served devices increases, since communication beams are more likely to illuminate uniformly the region. Furthermore, increasing the proportion of sensing APUs enhances sensing performance at the cost of reduced communication rates. Interestingly, for a given number of APUs, reassigning APUs to sensing roles improves sensing quality with no impact on the achievable rate, owing to the distributed nature of the architecture. Finally, we showed that sensing precision exhibits a non-monotonic dependence on the number of antennas, reflecting an interesting trade-off between array gain and illumination homogeneity across the service area.

\bibliographystyle{IEEEtran}
\bibliography{IEEEabrv,references}

\end{document}